# A Silicon Nanowire Ion-Sensitive Field-Effect-Transistor with elementary charge sensitivity

N. Clément[1], K. Nishiguchi[2], J.F. Dufreche[3], D. Guerin[1], A. Fujiwara[2], & D. Vuillaume[1]

(1) Institute of Electronics, Microelectronics and Nanotechnology, CNRS, University of Lille, Avenue Poincaré, BP60069, 59652, Villeneuve d'Ascq France

(2) NTT Basic Research Laboratories, 3-1, Morinosato Wakamiyia, Atsugi-shi, 243-0198 Japan

(3) CEA Marcoule, BP 17171, 30207 Bagnols sur Cèze cedex, France

**We investigate the mechanisms responsible for the low-frequency noise in liquid-gated nano-scale silicon nanowire field-effect transistors (SiNW-FETs) and show that the charge-noise level is lower than elementary charge. Our measurements also show that ionic strength of the surrounding electrolyte has a minimal effect on the overall noise. Dielectric polarization noise seems to be at the origin of the 1/f noise in our devices. The estimated spectral density of charge noise $S_q = 1.6 \times 10^{-2}$ e/Hz$^{1/2}$ at 10 Hz opens the door to metrological studies with these SiNW-FETs for the electrical detection of a small number of molecules.**

Semiconductor nanowire transistors are of great interest for future high performance[1-3] or low power electronics[4] and chemical sensors applications[5-9]. Evaluation of electrical noise is of prime interest to determine the performance limits of these devices. In particular, small transistors can reach elementary charge-sensitivity provided a very low noise level. This was first reached at very low temperature, and has led to considerable progress in solid-state physics including quantum-bit devices, counting statistics and spintronics. The high quality of Si/SiO$_2$ interface has allowed reaching elementary charge sensitivity, even at room temperature and has led to the fabrication of single-electron memories[4] or other single-charge based devices[10]. From a fundamental view point, the room-temperature elementary charge sensitivity opens the door to metrological studies that can only be done at room temperature such as electrical detection of ions or bio- and other functional molecules. However, it often requests an electrolytic environment and until now, liquid-gated transistors with elementary charge-sensitivity have not been demonstrated. Low-frequency noise in liquid-gated FETs often called Ion-sentitive or chemical FETs (ISFETs or CHEMFETs) was studied in conventional large size FETs[11] and more recently in carbon-nanotube[12] ISFETs. In both cases trapping-detrapping of many charges in the close channel proximity is the limiting low-frequency noise source. Here, we show that these two limiting noise source disappear using a 50 nm long Si nanowire covered with high quality thermal oxide and with the appropriate use of a back-gate.

A schematic view of the device is shown in Fig.1a. A droplet of deionized (DI) water with NaCl salt is inserted in a 0.86 mm² micro-bath made with a 300 µm-thick polymerized SU8 resist, Shipley©[13]. A platinum microelectrode is used to bias the droplet at voltage $V_g$ with SiNWs source electrode grounded. ISFETs usually use an Ag/AgCl electrode to bias the liquid. Here, we prefer to use a Pt microelectrode in our micro-bath because it is integrated, more stable and it does not modify the volume of the droplet. Moreover, we have confirmed that an Ag/AgCl reference electrode in the droplet measures the $V_g$ potential applied on the Pt microelectrode, except an offset, mainly due to differences in their work functions[13]. Optical Microscope image of the device (top view) is shown in Fig.1b. An atomic force microscope image of the SiNW is shown in Fig.1c. Large undoped silicon-on-insulator (SOI) channels are locally constricted and oxidized for the formation of an upper oxide with a thickness of 40 nm. The output drain current characteristics are determined by the constricted channel which width W and length L after the oxidation are 15 and 50 nm, respectively[14]. Such a small wire channel makes this SiNW-FET useful as a high-charge-sensitivity electrometer with single-electron resolution at room temperature[15-17]. Different possible sources of noise related to ISFETs are represented in Fig.1d: the Brownian motion of ions (ions induced noise), the fluctuation of dipoles in the oxide (DP noise), trapping-detrapping of an electron in a defect close to the SiNW (trapping noise) and thermal fluctuation of electrons in the channel (Johnson-Nyquist noise).

Fig.2a shows the output current vs. back-gate voltage (I-$V_{bg}$) reference curve (no droplet, device under a dry $N_2$ atmosphere), I-$V_{bg}$ for a floated (not biased) droplet of DI water and the output current vs. liquid top gate (I-$V_g$) curves for biased droplets of DI water with NaCl at different ionic strengths ($10^{-4}$ M to 1M). Fits (solid lines) are obtained with the classical MOSFETs equations in the linear regime with a transverse field-dependent mobility, which allows us to extract the threshold voltage[13]. Fig. 2b is a zoom of Fig.2a at low gate voltages to focus on biased droplet at different ionic strengths. The threshold voltage $V_{th}$ follows a decrease of about 60 mV/dec vs. (pNa=log[$Na^+$]) according to Nernst law which gives the dependence on the concentration of $Na^+$ ([$Na^+$]) (Fig. 2b, inset). As proposed initially by Bergveld[18], ISFET's are sensitive to Nernst potential that depends on the concentration of ions having affinity with the gate insulator surface. As a consequence, increasing the concentration of such ions can shift the threshold voltage of the liquid-gated SiNW-FET up to 59 mV/dec. For nanowire ISFETs, sensitivity to pH has been previously shown[7]. Here we show that other ions such as $Na^+$ can be sensed with such devices. Capacitance, mobility and subthreshold swing S ≈ 350 mV/dec barely depends on ionic strength. This is mainly because the dielectric constant of $H_2O$ is much larger than that of $SiO_2$ and therefore the electrolytic double layer capacitance does not affect the gate capacitance.

We have measured the low frequency power spectrum current noise $S_I$ following a protocol described elsewhere[13,19]. Fig.3a shows $S_I$ at $V_{bg}$ = 12 V for the reference (no electrolyte) device and Fig. 3b shows $S_I$ at different gate voltages $V_g$ for the biased droplet with [NaCl] = 10 mM. We see in both cases that $S_I$ scales as 1/f at low frequency and becomes constant (white noise) above a corner frequency which can be as low as 70 Hz due to the very low 1/f noise level in our SiNW transistor. 1/f noise amplitude increases first with $V_g$ and then decreases, which is not the case for the thermal white noise at higher frequency as illustrated in Fig. 3c. This is because $S_I$ at 10 Hz scales as $g_m^2$ and not $I^2$, where $g_m=\partial I/\partial V_g$ is the transconductance (black line in Fig.3c). The noise at 3 kHz (red line in Fig.3c) is well fitted with a SiNW thermal noise $S_I = 4kTV_d/I$ where k is the Boltzman constant, T the temperature, $V_d$ the drain voltage and I the average drain current. Fig. 3d shows the power spectrum voltage noise $S_v = S_I/g_m^2$ (i.e. the power spectrum noise referred to the input gate) at 10 Hz as a function of $V_{bg}$ and $V_g$ for the reference sample (no droplet) and the liquid-gated sample, respectively. We note that Sv is constant with gate voltage and there is an order of magnitude of difference in noise between the reference and liquid-gated sample. The behavior and origin of this 1/f noise will be discussed later.

Following a careful and detailed calibration procedure reported elsewhere[20], we obtained, for this device, $C_g$ = 6.25 aF for the upper gate capacitance (liquid-gated) and $C_{bg}$ = 0.66 aF for the back-gate capacitance (reference sample). From $S_v$, we can estimate the charge noise at 10 Hz $Sq_{bg}^{1/2}$ = 7x10$^{-3}$ e/Hz$^{1/2}$ for reference sample and $Sq_g^{1/2}$=1.6x10$^{-2}$ e/Hz$^{1/2}$ for the liquid-gated configuration using

$$Sq_g^{1/2} = \frac{C_g}{q} S_{vg}^{1/2}; \quad Sq_{bg}^{1/2} = \frac{C_{bg}}{q} S_{vbg}^{1/2} \qquad (1)$$

This charge noise level is low enough to get elementary charge sensitivity at room temperature. As a comparison, at very low temperature, charge noise level of $10^{-3}$ e/Hz$^{1/2}$ at 10 Hz for charge-sensitive electrometers is a standard value[21,22]. Another easy way to confirm the elementary charge sensitivity is the observation of a discrete two levels fluctuation (RTS) due to the trapping/detrapping of a single electron in a defect located in the oxide[13].

The 1/f noise in ISFETs could have several origins including trapping/detrapping of electrons by a huge number of oxide defects close to electrons in the channel, noise due to mobility fluctuations, ion-induced noise (Brownian motion of ions), or dielectric polarization noise[23,24] (DP noise).

In the case of mobility noise, the power spectrum current noise $S_I$ should scale with $I^2$ and not with $g_m^2$ as shown in Fig. 3c (black curve). Ion-induced noise is also negligible in our devices because $S_v$ does not depend on ionic strength (Fig. 3d) and the measured white noise follows thermal noise of the SiNW (Johnson-Nyquist noise, Fig. 3d, red curve). Theoretical derivation of ion-induced noise will be reported elsewhere[25]. It confirms the lower contribution of ion induced noise. Although we can observe trapping-detrapping noise via two level fluctuations of current, the so-called RTS[13], traps are in the worst case active one by one in turn with gate voltage[26] and in the better case there are no traps in our devices. RTS amplitude in our 50 nm long SiNW does not depend on trap's position along the SiNW or on the trap depth in the oxide surrounding the SiNW[26]. This feature prohibits that the 1/f noise is a sum of many trapping-detrapping events, each one leading to Lorentzian noise[26]. Finally, for trapping-detrapping noise[26,11] also called at low temperature background charge noise[21], $S_v$ scales as $1/C^2$ and not as $1/C$ as shown in Fig. 3d. Indeed, in Fig. 3d, there is about one order of magnitude between $S_v$ for the reference SiNW and $S_v$ for the liquid-gated SiNW but estimated capacitance have only a factor 10 in difference. On the opposite, DP noise is a good candidate for the observed 1/f noise in our device. Eq.2 is derived from the fluctuation-dissipation theorem which extends Johnson-Nyquist noise equation to complex impedances[23,27].

$$S_{vg} = 4kT \, \text{Re}(Z) = \frac{2kT \, tg\delta}{\pi C_g f}; \quad S_{vbg} = \frac{2kT \, tg\delta}{\pi C_{bg} f} \quad (2)$$

where $tg \, \delta = \varepsilon''/\varepsilon'$ is the dielectric loss tangent with $\varepsilon'$ and $\varepsilon''$ the real and imaginary dielectric permittivities of the oxide, respectively. $SiO_2$ is considered as a low loss / flat loss dielectric[28]. A standard value for high-quality thermally grown oxide[28] is $tg \, \delta \approx 3.8 \times 10^{-3}$. Considering $C_g = 6.25$ aF and $C_{bg} = 0.66$ aF as discussed previously, we get using eq. 2 and the $S_{Vg}$ and $S_{Vbg}$ in Fig.3d, $tg \, \delta = 7.8 \times 10^{-3}$ for back oxide and $4.3 \times 10^{-3}$ for front oxide, which are in the range of expected results.

This study is very usefull to investigate the best structure for optimum ISFET's sensitivity. For example, let us consider a single molecule detection, for instance a single redox event (single charge detection), the associated current variation is $\Delta I = g_m \cdot \Gamma \cdot q/C_g$. $\Gamma$ is a parameter to account for molecule screening effects[8]. ($\Gamma=1$, for no screening, $<1$ otherwise). The signal to noise ratio in our ultra-low noise SiNW-FET can be expressed as:

$$\Delta I / SI^{0.5} = \frac{\Gamma.q.(\pi f)^{0.5}}{(2kT.tg\delta.C_g)^{0.5}} \qquad (3)$$

The dielectric loss should be as low as possible. Therefore, native oxides or not optimized oxides should be avoided. Finally $C_g$ should be as small as possible which implies nanometric dimensions for the SiNW.

To conclude, we have demonstrated that liquid-gated SiNW ISFETs can have an elementary charge sensitivity ($S_q$ = 1.6x10$^{-2}$ e/Hz$^{1/2}$ at 10 Hz) making it a suitable device for single molecule detection with a signal to noise ratio larger than 50. Although the threshold voltage of this device is sensitive to NaCl concentration, we show that ion-induced noise is negligible in the device. Among the several possible sources of 1/f noise in this device, only the dielectric polarization noise gives a quantitative agreement with the experimental data. These results open the door to metrological studies with SiNW-FET for the electrical detection of a small number of molecules.

Acknowledgements:
The authors would like to thank S. Ascott for helpful advises for the fabrication of 500 µm-thick micro-baths fabrication.

CAPTIONS.

Fig. 1: a) 3D schematic view of the device. The Si s ubstrate can act as a back-gate biased at $V_{bg}$. SiNW source is grounded ($V_s$ = 0V) and drain is biased at $V_d$. A micro-bath is fabricated with polymerized SU8 resist, Shipley©. Droplet voltage is fixed using a platinum electrode inserted in the micro-bath (voltage $V_g$). SiNWs are surrounded by $SiO_2$. b) Optical image of the experimental setup. Needles bias the device from outside the micro-bath which is necessary to avoid any leakage current and pollution of the droplet. c) AFM image of the SiNW. d) Schematic cut view of the SiNW that explains the different sources of noise: ions induced noise, DP noise, trapping noise and thermal noise.

Fig. 2: a) I-$V_{bg}$ curve for the reference sample (no electrolyte), with electrolyte in the µ-bath (floated) and liquid-gated with different NaCl concentrations. Since I-$V_g$ curves for different NaCl concentrations are tight, a zoom is shown in b). Inset) Threshold voltage is shifted according to Nernst law on pNa. DI in the scale bar means DI water (no NaCl).

Fig. 3:a) Example of power spectrum current noise for the reference when $V_{bg}$ = 12 V. It follows 1/f law at low frequency and above a corner frequency the Johnson-Nyquist noise dominates. b) Liquid-gated power-spectra current noise for different Vg (concentration of NaCl = 10 mM). c) Gate-voltage dependence of $S_I$ at 10 Hz (1/f noise) α $g_m^2$ and white noise at 3 kHz ≈ 4kTI/$V_d$. d) $S_V=S_I/g_m^2$ at 10 Hz as a function of $V_g$ for different ionic strengths and as a function of $V_{bg}$ for reference. The decrease of $S_v$ by about an order of magnitude when the SiNW is liquid-gated is in accordance with DP noise (Eq. 2).

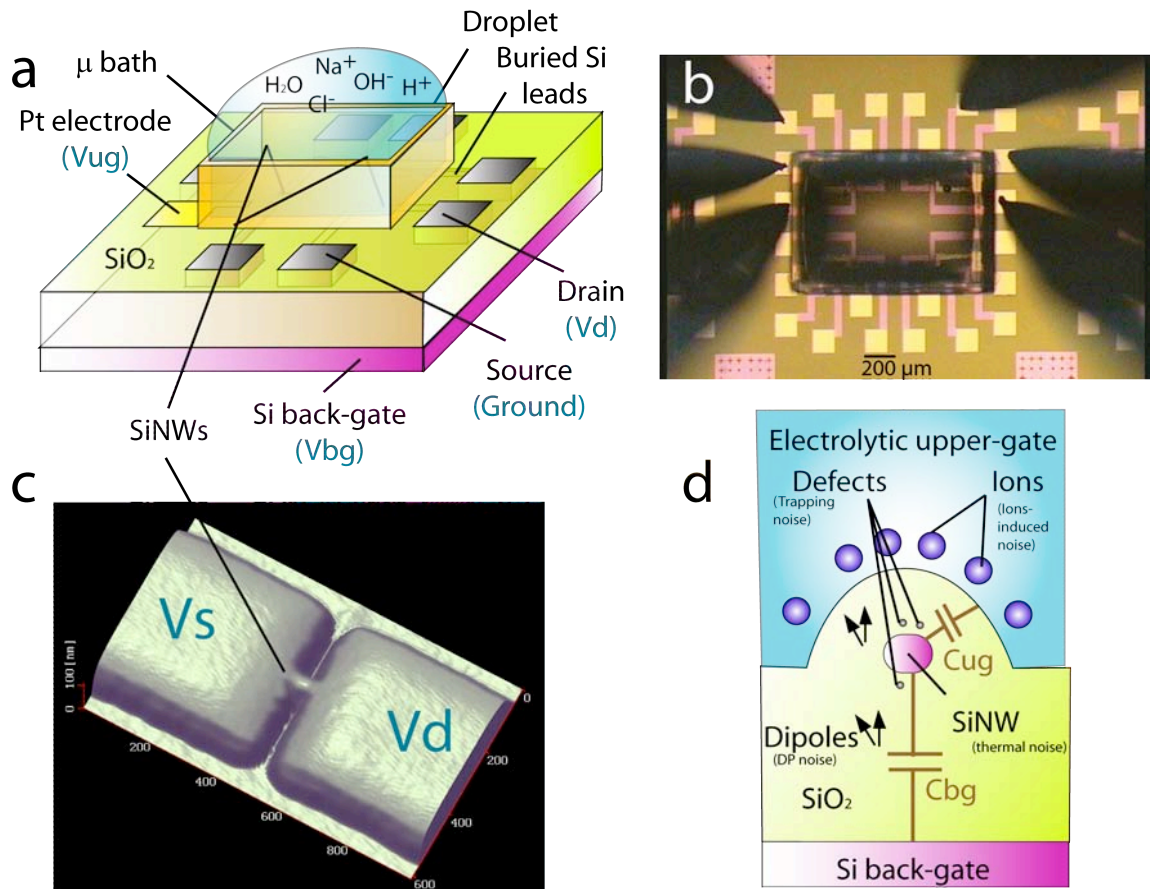

Fig.1

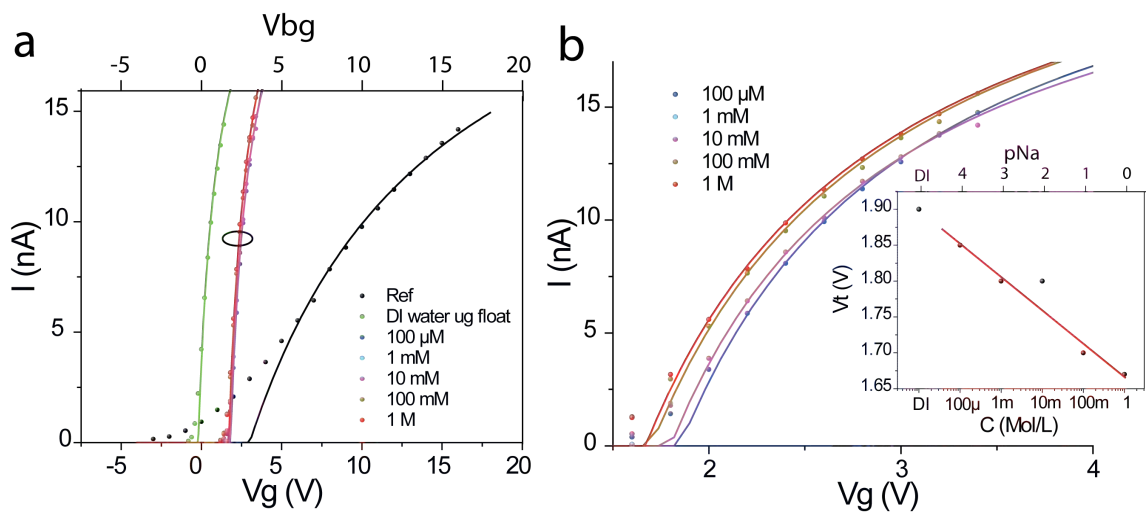

Fig 2

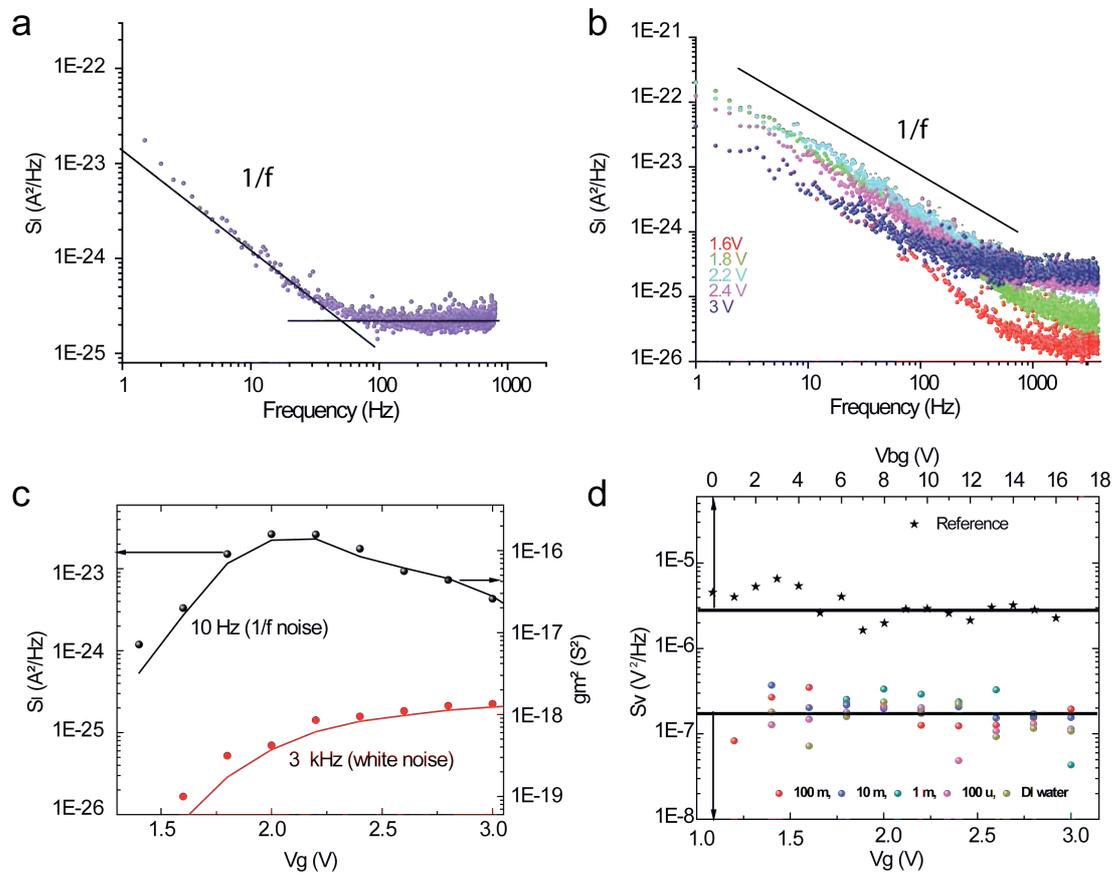

Fig.3

# A Silicon Nanowire Ion-Sensitive Field-Effect-Transistor with elementary charge sensitivity

N. Clément[1], K.Nishiguchi[2], J.F. Dufreche[3], D. Guerin, A.Fujiwara[2], & D. Vuillaume[1]

(1) Institute of Electronics, Microelectronics and Nanotechnology, CNRS, Avenue Poincaré, 59652, Villeneuve d'Ascq France

(2) NTT Basic Research Laboratories, 3-1, Morinosato Wakamiyia, Atsugi-shi, 243-0198 Japan

(3) CEA Marcoule, BP 17171, 30207 Bagnols sur Cèze cedex, France

**Supplementary information**

*Section 1: Fabrication of the Pt microelectrode and SU8 micro-bath*

Pt microelectrodes (100 nm) are made with conventional optical lithography techniques. The fabrication of the 300 µm – thick micro-bath is more complex. We use two layers of SU8 resist (SU8 2002, 2 µm and SU8 2075, 500 µm). The first layer is used to get a good adhesion to the substrate. Annealing should be made with a temperature ramp to avoid only surface baking. Then optical alignment process should be made in proximity mode because at this stage resist is not completely hard. After development, we make a hard baking at 190°C for two hours to polymerize the resist. An example of empty micro-bath is shown in Fig. S1a and filed in Fig.S1b.

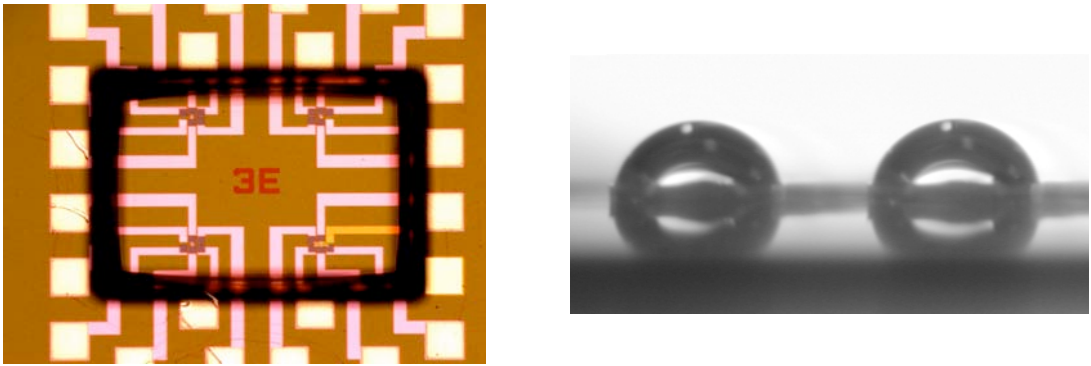

Fig.S1 (a) Optical image of 300 µm-thick micro-bath made with SU8 resist. (b) Side photography of micro-droplets inserted in micro-baths.

## Section 2: Calibration with an Ag/AgCl reference electrode

We have compared the droplet potential applied by the Pt electrode with an Ag/AgCl reference electrode. A schematic view of the experimental setup is shown in Fig. S2a. It is the same as the one in Fig. 1a, except the electrode added to measure the droplet voltage $V_m$ between the droplet and the ground with Ag/AgCl potential reference. We find that as we linearly increase $V_g$ (Platinum electrode), $V_m$ follows $V_g$ with an initial offset of 0.503 V (Fig.S2c). This value is given as information, but in the article, we refer to the potential applied on the Platinum electrode. The home-made reference micro-electrode has been fabricated following a protocol described in ref [1]. An optical microscope image of the reference electrode is shown in Fig.S2b.

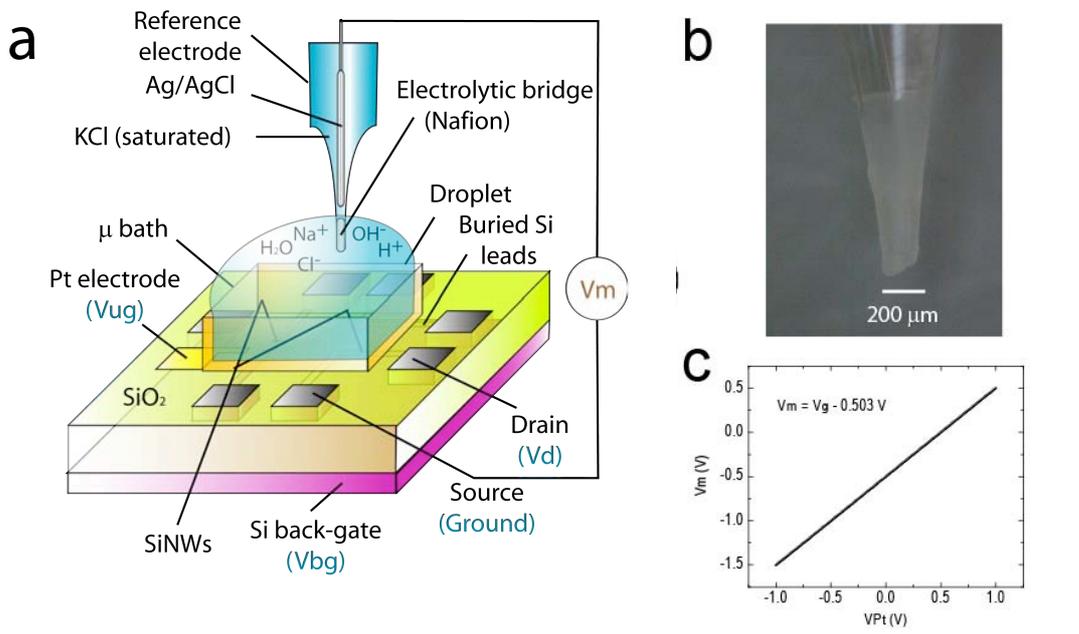

Fig.S2: Schematic view of the ISFET structure with an Ag/AgCl reference electrode (a), optical image of the reference electrode (b) and $V_m$ measured as $V_g$ is swept from -1 to +1 V.

### Section 3: Fit of I-Vg curves with conventional MOSFET equations

To fit the curves in Figs. 2, we used the standard equation of MOSFET.

$$Id = \frac{\mu C_g}{L^2}(Vg - Vth)Vd \quad \text{with} \quad \mu = \frac{\mu_0}{1 + \theta^{-1}(Vg - Vth)} \quad \text{(Eq. S1)}$$

where μ is the mobility, $\mu_0$ the mobility at low transverse field and Θ is a constant to account for the reduction of mobility with transverse electric field. $V_{th}$ is the threshold voltage, $V_d$ = 1mV the drain voltage as shown in Fig.1a. $\mu_0$ at room temperature in our devices was found to be relatively constant from device to device. Following a careful calibration procedure with 4 different techniques to estimate capacitances detailed elsewhere[2], we obtained, for this device, $C_g$ = 6.25 aF for the upper gate capacitance (liquid-gated) and $C_{bg}$ = 0.66 aF for the back-gate capacitance (reference sample). Fits with eq. S1 in Figs. 2a and 2b give $\mu_0$ = 64 cm$^2$/V.s and Θ=12 V$^{-1}$, $\mu_0$ and $C_g$ being independent on ions concentration. Fig. S3 is Fig.2b plotted in log-scale.

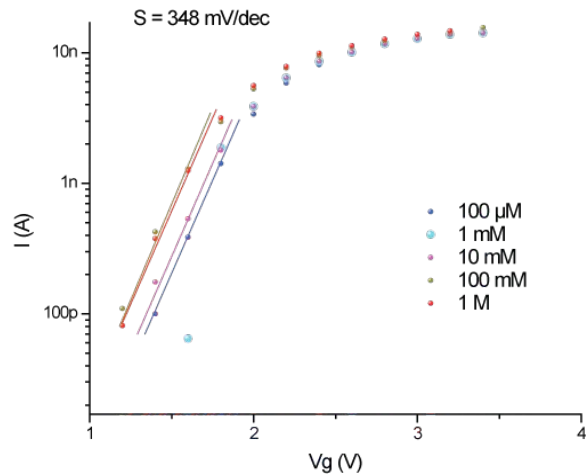

Fig.S3: I-V$_g$ curves in log scale for different ionic strength. The subthreshold swing does not depend on ionic strength.

## Section 4: 50 Hz and harmonics peaks

50 Hz peaks and harmonics come from extrinsic noise (grounds). Since it has been removed in Figs 3a, 3b, we show here raw curves in Fig. S4.

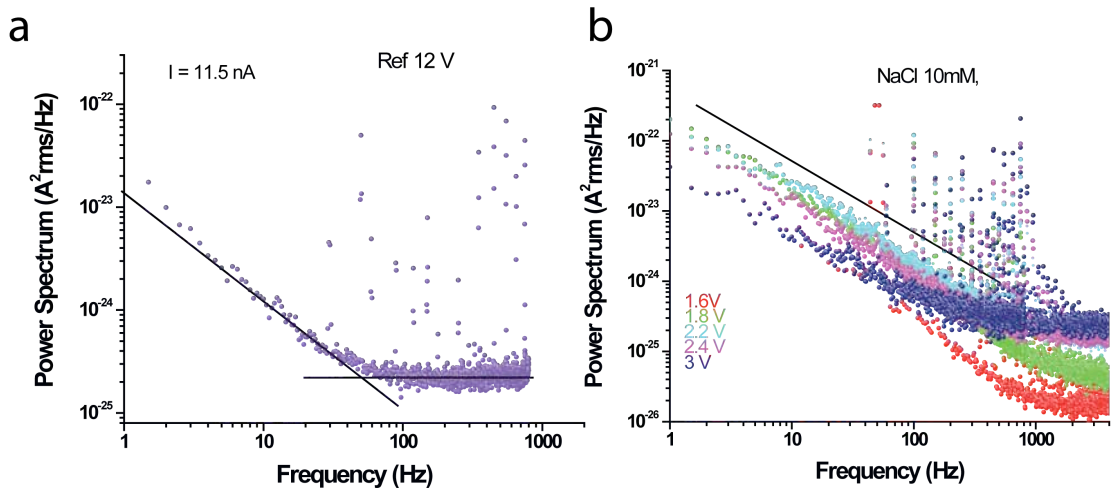

Fig.S4: same legends as Figs 3a, 3b, 4e

## Section 5: Random Telegraph Signal due to the trapping/detrapping of a single electron near the SiNW channel

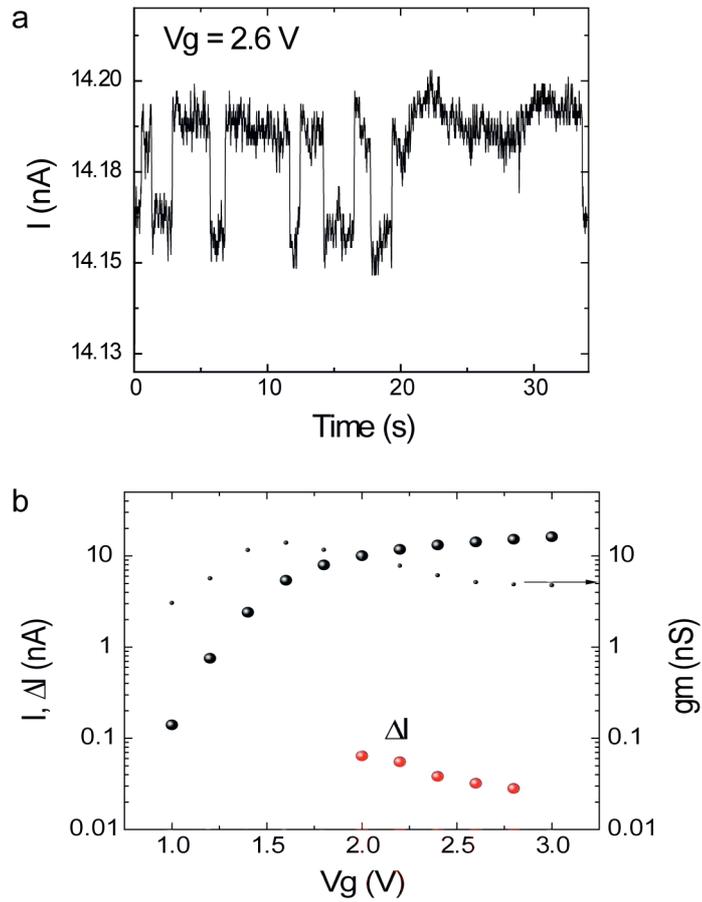

Fig. S5a is an example of Random-Telegraph-Signal (RTS), a stochastic two-level fluctuation of current (amplitude ΔI). Fig. S4b shows I-$V_g$, ΔI-$V_g$ and $g_m$-$V_g$. ΔI is proportional to $g_m$ with ΔI/$g_m$ ≈ 0.37 q/$C_g$.[2]

[1] T. Kitade, K. Kitamura, S. Takegami, Y. Miyata, N. Nagatono, T. Sakaguchi and M. Furukawa, Anal.Sci. 21, 907 (2005)

[2] N. Clement, K. Nishiguchi, A. Fujiwara and D. Vuillaume, IEEE Trans. on Nanotech., submitted.